\documentclass[twocolumn,prb,superscriptaddress]{revtex4-1}
\usepackage{dcolumn,amsfonts,latexsym,amsmath,amsthm,amssymb,amsmath,amscd,graphicx,mathrsfs,tikz-cd,physics,braket,hyperref}

\newcommand{\p}{\partial}

\newcommand{\mH}{\mathcal{H}}
\newcommand{\sig}{\sigma}

\newcommand{\bk}{\mathbf{k}}

\newcommand{\bq}{\mathbf{q}}
\newcommand{\br}{\mathbf{r}}

\newcommand{\bx}{\mathbf{x}}

\newcommand{\sgn}{\text{sgn}}

\begin{document}

\title{ 
Topological magnetic textures in magnetic topological insulators}
\author{ Nisarga Paul}
\affiliation{Department of Physics, Harvard University, Cambridge, MA, USA}
\affiliation{Department of Physics, Massachusetts Institute of Technology, Cambridge, MA,
USA}
\author{Liang Fu}

\affiliation{Department of Physics, Massachusetts Institute of Technology, Cambridge, MA,
USA}

\begin{abstract}
The surfaces of intrinsic magnetic topological insulators (TIs) host magnetic moments exchange-coupled to Dirac electrons. We study the magnetic phases arising from tuning the electron density using variational and exact diagonalization approaches. In the dilute limit, we find that magnetic skrymions are formed which bind to electrons leading to a skyrmion Wigner crystal phase while at higher densities spin spirals accompanied by chiral 1d channels of electrons are formed. The binding of electrons to textures raises the possibility of manipulating textures with electrostatic gating. We determine the phase diagram capturing the competition of intrinsic spin-spin interactions and carrier density and comment on the possible application to experiments in magnetic TIs and spintronic devices such as skyrmion-based memory. 
\end{abstract}
\maketitle

\section{Introduction}

Topological materials, topological excitations and topological quantum effects have become a very active research frontier in condensed matter physics. Manifestations of topology in momentum space include the integer quantum Hall effect\cite{TKNN}, Chern insulators\cite{HaldaneModel}, and topological insulators\cite{KaneMele2005, HasanKane2010} (TIs). In real space, topological order parameter configurations such as vortices, monopoles, and skyrmions are ubiquitous in spin systems and magnetism\cite{KosterlitzThouless1973,Castelnovo2008,Robler1973,Muhlbauer2009,nagaosa3,Balents2017,Ochoa2019}. In particular, magnetic skyrmions have been the focus of much recent study, spurred by their observation in room-temperature materials\cite{Yu2012,Woo2016,Lin2018}, stability\cite{Jiang2015,Buttner2018,bogdanov2}, and suitability in spintronics for next-generation memory devices \cite{Fert2017}. \par

Both momentum space and real space topology can play crucial roles in magnetic topological insulators, where Dirac electrons interact with topological spin textures associated with magnetic moments\cite{PhysRevB.82.161401,hurst}. Recently the stoichiometric compound MnBi$_2$Te$_4$, a TI containing a periodic sublattice of Mn$^{2+}$ ions with magnetic moment $5\mu_B$, was synthesized and studied for the first time\cite{otrokov2019, gong2019}, opening the new field of intrinsically magnetic TIs. The compound comprises alternating quintuple layers of the  topological insulator Bi$_2$Te$_3$ and ferromagnetic (FM) MnTe bilayers. A single septuple layer (SL) thin film exhibits ferromagnetic order.  
The exchange coupling between FM-ordered magnetic moments and Dirac surface electrons on the adjacent TI layer can open up a gap at the Dirac point and give rise to nontrivial topology in momentum space, characterized by quantized Chern number\cite{FuKane2007}.  The resulting Chern insulator and zero-field quantum anomalous Hall state have recently been observed in transport measurements on exfoliated few-layer Mn-Bi-Te flakes.  

Unlike TIs with randomly doped magnetic impurities, intrinsic magnetic TIs are stoichiometric compounds containing a lattice of magnetic atoms coupled to topological electrons. This feature not only promises a  magnetic ordering and quantum anomalous Hall effect at temperatures as high as $50$ K\cite{2011.07052,Lei27224}, but also opens the exciting possibility of magnetic control of topological electronic properties and electrical control of magnetic order. In particular, magnetic textures such as spirals and skyrmions can arise  at surfaces and interfaces from the chiral Dzyaloshinskii-Moriya (DM) interaction \cite{Moriya1960,Dzyaloshinskii1957,Dzyaloshinskii1963} due to broken inversion symmetry. Magnetic TI surfaces  provide a platform of this kind. This motivates us to consider the effect of real-space magnetic structures on topological Dirac electrons, which may enable the manipulating magnetic domains and textures by electric currents and electrostatic gating.

\par

\par

\begin{figure*}
\includegraphics[width=0.9\linewidth]{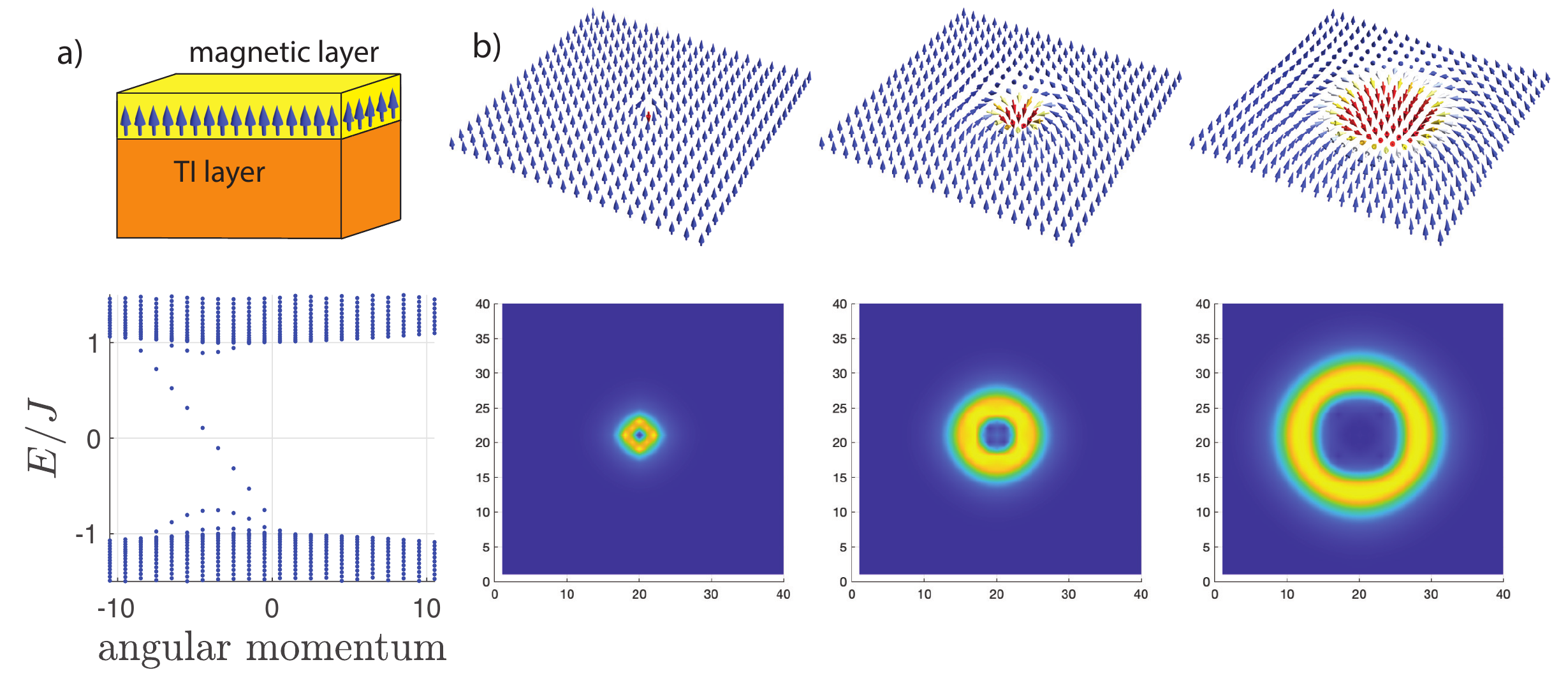}
\caption{(a) Top, illustration of an intrinsic magnetic TI. In the case of MnBi$_2$Te$_4$ the alternating layers consist of the TI Bi$_2$Te$_3$ and magnetic layer MnTe. The magnetic layer can host real-space topological textures such as N\'eel skyrmions, right. Below, spectrum of Dirac electrons coupled to N\'eel skyrmion texture with radius $b=10$ lattice spacings and shape parameters $\alpha=3/2,\beta=1/2$  (Eq. \eqref{eq:w}) plotted against total (spin + orbital) angular momentum, showing a branch of chiral mid-gap bound states. (b) Top, N\'eel skyrmion textures with radii $b=1,5,$ and $10$ lattice spacings and same shape parameters. Below, magnitude of spin up component of the smallest $|E|$ Dirac bound state in the presence of the corresponding texture.}
\label{fig:sk}
\end{figure*}

Motivated by recent advances in magnetic TIs and skyrmion physics, we study the magnetic phenomena arising from doping Dirac electrons. We focus separately on the cases of (a) a single added electron, (b) low carrier density and (c) high carrier density. We predict the formation of skyrmion textures with localized Dirac modes for (a) and (b) and the formation of stripes with 1d channels of chiral modes for (c). Because the ground state magnetic order parameter depends on carrier density, we obtain electrically tunable skyrmion and stripe phases whose periods vary with density.
\par

As a potential application, we envision that intrinsic magnetic TIs can open the path for a memory device storing information using magnetic skyrmions in a low-carrier density material. The use of skyrmions for information storage has been reported in magnetic metals, where the exchange interaction couples itinerant electrons and localized spins. However, due to the large carrier density of metals,  a large current density is required to drive skyrmions in writing and reading out data, which results in Joule heating and considerable power consumption. This drawback can be alleviated by working with skyrmions in a low-carrier density, bulk insulating material such as a magnetic TI. As we shall show, a skyrmion in a magnetic TI carries tightly bound electric charge. Under the right conditions, these charged skrymions are the only charge carrier at low doping. Therefore, a relatively small current is sufficient to drive the skyrmion motion.

\par

\textit{Model.} For the Hamiltonian of the 2d surface of a magnetic TI describing Dirac electrons coupled to a dense array of $N$ classical spins $\vec S=(S^x,S^y,S^z)$ we take

\begin{equation}
H =  H_e +H_{eS}+H_S
\end{equation}
where the first two terms,

\begin{equation}\label{eq:dirac}
H_e=v_F \sum_\bk \hat c_\bk^\dagger(k_x \sig^y-k_y\sig^x-E_F) \hat c_\bk
\end{equation}
and
\begin{equation}\label{eq:ex}
H_{eS} =  -\frac{J}{\sqrt{N}}\sum_{\bk,\bq} \hat c_{\bk+\bq}^\dagger \vec S_\bq \cdot \vec \sigma \hat c_\bk,
\end{equation}

are a Rashba Hamiltonian for massless Dirac electrons and an exchange interaction, respectively. The last term consists of intrinsic spin interactions, including a ferromagnetic exchange interaction, DM interaction, Zeeman field, and out-of-plane anisotropy, namely
\begin{multline}\label{eq:LG}
H_S = -A \sum_{\br,i} \vec S_\br \cdot \vec S_{\br+\mathbf{e}_i}+ D \sum_{\br,i}[ \mathbf{e}_i \times (\vec S_\br \times \vec S_{\br+\mathbf{e}_i})]_z\\
- \vec B\cdot \sum_\br \vec S_\br -K\sum_\br S_{z\br}^2
\end{multline}
where $z$ is the out-of-plane direction. We have chosen a normalization so that $S_\br^2=1$. The form of the DM interaction above is appropriate for a lattice with $C_{nv}$ symmetry, favoring the formation of spirals rotating \textit{along} the wavevector (N\'eel type), as opposed to orthogonal to the wavevector (Bloch type).

The phase diagram of the magnetic moments \textit{without} coupling to electrons has been studied both theoretically\cite{bogdanov1, bogdanov2} and experimentally \cite{Muhlbauer2009,Nagaosa2013,Yu2011,Onose2012}. With vanishing Zeeman field and anisotropy, the ground state of $H_S$ is a periodically modulated spiral texture with wavevector $q\sim D/A$ and N\'eel wall-like rotation. With increasing anisotropy, the spiral degenerates into a system of domain walls which is energetically favorable over the uniform state when $\sqrt{D^2/AK}$ is above a threshold. Turning on an out-of-plane Zeeman field penalizes the large areas of antiparallel spins, resulting in a phase transition to a skyrmion crystal (SkX) above a critical field and to a uniform state above a higher critical field. As the temperature is increased the spiral or SkX order is destroyed.
\par

We now consider the exchange coupling between localized spins and topological surface electrons. As we shall show below, the interplay between magnetic and electronic degrees of freedom leads to (1) a carrier density dependent magnetic phase diagram; and (2) chiral electronic states. Furthermore the exchange coupling is capable of driving the formation of charged skyrmions at small electron doping and Coulomb repulsion is capable of leading to a skyrmion Wigner crystal (SWX).

The remainder of this paper is organized as follows. In Sec. \ref{sec:skyrmion} we study the formation of skyrmion textures from doping a single electron and the formation of a SWX for a very dilute density of electrons. In Sec. \ref{sec:stripe} we study the stripe phase which forms at finite density using exact diagonalization and discuss the 1d channels of chiral modes bound to the stripes. We then present a phase diagram of competing stripe orders when $H_{eS}$ and $H_S$ are both important, and in Sec. \ref{sec:end} we conclude.

\section{Skyrmion formation}
\label{sec:skyrmion}

\begin{figure}
  \centering
  \begin{minipage}[b]{0.24\textwidth}
    \includegraphics[width=\textwidth]{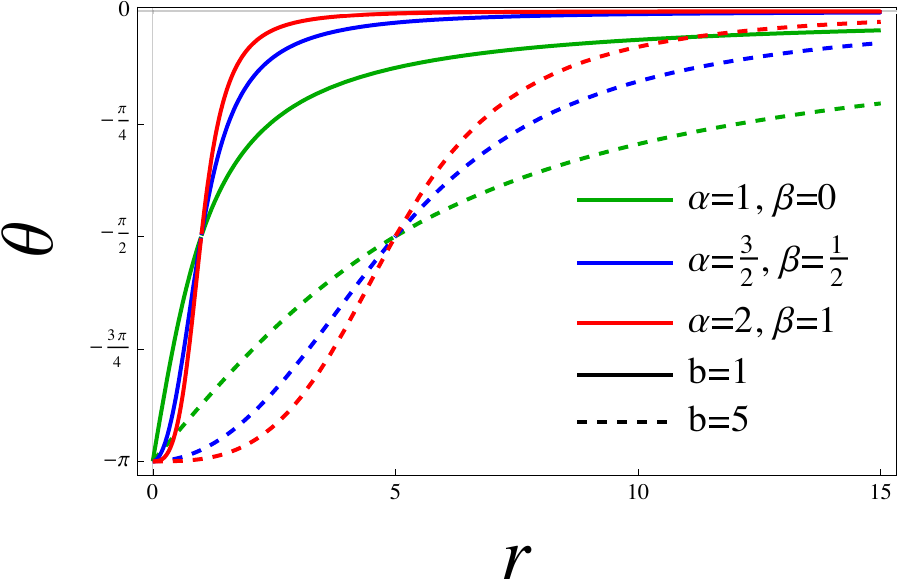}
  \end{minipage}
  \begin{minipage}[b]{0.23\textwidth}
    \includegraphics[width=\textwidth]{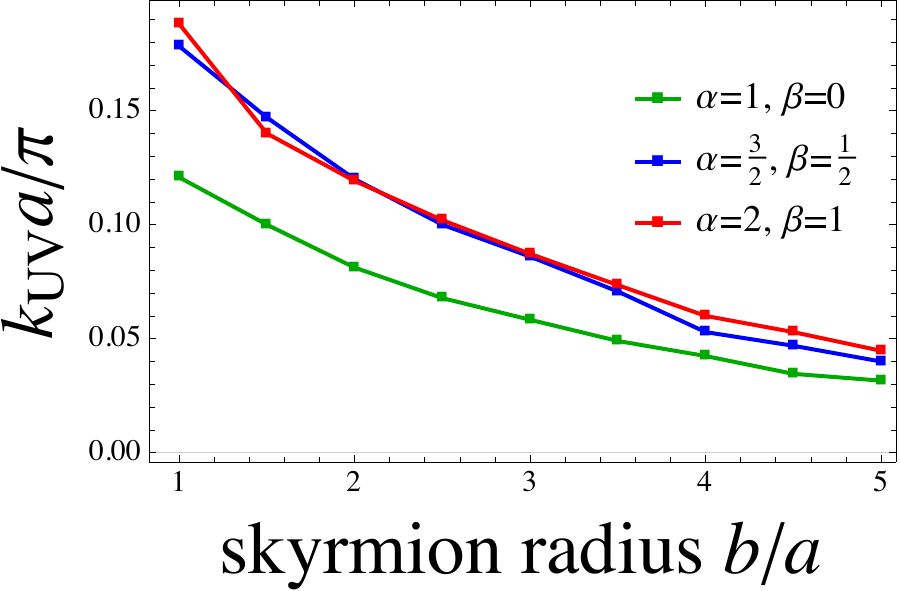}
  \end{minipage}
  \caption{Left, skyrmion profiles $\theta(r)$ where $\theta=0$ corresponds to $+\hat z$, for the shape parameters shown (c.f. Eq. \eqref{eq:w}). Right, the cutoff $k_{UV}$ below which a skyrmion of the given radius becomes favorable upon doping a single electron. (Distances measured in lattice spacings.)}
  \label{fig:sk2}
\end{figure}

Eq. \eqref{eq:LG} has a uniform ground state $\vec S_\br = \pm \hat z$ when the direct exchange interaction dominates, and the coupled electrons form two bands at energies $\pm \sqrt{J^2+v_F^2|\bk|^2}$. At charge neutrality, the lower band is filled and the addition of another electron incurs an energy cost of $2J$ unless a nontrivial texture forms. Eq. \eqref{eq:LG} is known to support skyrmions as excitations, so we investigate the electron spectrum in the presence of a skyrmion. We plot this spectrum in Fig. \ref{fig:sk}a. We observe mid-gap bound states, indicating the possibility that the system will spontaneously form a skyrmion to accommodate the extra electron and a skyrmion Wigner crystal for a very dilute density of electrons (Fig. \ref{fig:sk}b).
\par 
Let us work in the strong $J$ limit, neglecting energy costs associated to $H_S$. As a variational approach, we consider a 50$\times$50 square lattice hosting a skyrmion texture and add a single electron to the charge-neutral state. Then we compare the total energy to that of the uniform state with a single added electron. We take a tight-binding approximation for $H_e$ suitable for the lattice, which we describe in App. \ref{app:diag}; it has single Dirac cone at $\bk = 0$. We choose a momentum cutoff $|\bk|< k_{UV}$ to indicate the region of linear dispersion. We parameterize the skyrmion profiles, shown in Fig. \ref{fig:sk2}, using the following ansantz: 
\begin{multline}\label{eq:w}
(S^x,S^y,S^z)= \frac{(w+\bar w,-i(w-\bar w),1-|w|^2)}{1+|w|^2},\\
 w(x, y) = e^{i\gamma}\frac{b^{\alpha+\beta}}{\bar z^\alpha z^\beta},\quad z= x+iy.
\end{multline}
To our knowledge this is a new parameterization, which was chosen for its compact description of a four-parameter family of skyrmion profiles: $\gamma$ adjusts in-plane spin orientation (we set $\gamma=\pi$ for a N\'eel-type skyrmion), $b$ sets the radius, $\alpha-\beta$ is the vorticity, and $\alpha+\beta$ defines the sharpness of the domain wall. For example, a hard-wall magnetic bubble of radius $b$ is recovered by taking $\alpha=\beta \rightarrow \infty $. Indeed, any smooth magnetic texture can be captured by some smooth function $w(x, y)$. Skyrmion textures resulting from holomorphic $w$, a special case, were discussed in Ref. [\onlinecite{kuchkin}] and a one-parameter family of skyrmions\cite{PhysRevB.82.094429} is recovered by taking $\alpha=0,\beta=-1,\gamma=\pi/2$. Higher-winding skyrmions, multiple skyrmions, or even a skyrmion crystal can be achieved by taking appropriate sums of the above ansatz. Such a sum may be preferable to the typical sinusoidal ansatz when the core size of the skyrmions is uncorrelated with their spacing, as in the skyrmion Wigner crystal we discuss shortly.

\par
In Fig. \ref{fig:sk2} we show that the system prefers to form a N\'eel skyrmion (with vorticity 1) to accommodate the extra electron assuming a cutoff $k_{UV}a$ consistent with a Dirac cone dispersion, where $a$ is the lattice spacing. A Bloch skyrmion would be favored for a non-Rashba spin-orbit coupling. The added electron binds to the skyrmion with profiles we numerically determined in Fig. \ref{fig:sk}. The radius $b$ of the skyrmion satisfies $a \lesssim b \lesssim 1/k_{UV}$. The cutoff dependence indicates that in real materials for which the Dirac cone is a low-energy approximation, the full dispersion may be relevant in determining skyrmion size. For instance from ARPES data\cite{otrokov2019} for MnBi$_2$Te$_4$ the surface states' Dirac cone remains a good low energy approximation up to $k_{UV}\sim .05$ \AA$^{-1}$, beyond which the bulk bands are important. Since $a \approx 4$ \AA\, our numerics would yield a skyrmion texture of $b/a \sim 3$, an extended object comprising $\sim 10$ Mn atoms. We note that although using the family of textures in Eq. \eqref{eq:w} shows that in principle a skyrmion forms we have not optimized over all skyrmion shapes. Moreover, for real materials the direct spin-spin interactions, including the ferromagnetic exchange and DM interaction, may also help to stabilize skyrmions and affect their size.

Let us approximate the spectrum of Dirac modes in the presence of a skyrmion. Working in polar coordinates, we consider the skyrmion of radius $b$

\begin{equation}
(S^r,S^\theta,S^z) = \begin{cases} (0,0,-1) & r<b\\ (-1,0,0) &b<r<b+\delta\\
(0,0,1) &r>b+\delta
\end{cases}
\end{equation}
with $\delta \ll b$. This is an idealized approximation to the skyrmions shown in Fig. \ref{fig:sk}. Dirac bound states are localized near $b$, with a half integer angular momenta $m$ satisfying $|m|v_F/b\lesssim J$. We solve the continuum Dirac equation in detail in Appendix \ref{app:skyrm}. We find mid-gap energies

\begin{equation}\label{eq:skyrmspec}
E \approx mv_F/b - \delta J^2f(Jb/v_F)/v_F
\end{equation}
where $f$ is a bounded, positive function of the skyrmion size whose explicit form is given in Eq. \eqref{eq:fr0}. Thus small in-plane component $\delta>0$ results in a downward spectral shift, breaking particle-hole symmetry. This provides another way to understand why skyrmion formation is favorable. \par

The binding of Dirac modes to skyrmions was previously studied in Refs. [\onlinecite{PhysRevB.82.161401}] and [\onlinecite{hurst}]. Ref. [\onlinecite{PhysRevB.82.161401}] studied the general relations between magnetic textures and their induced electric charge, as well as their motion in an electric field (see also [\onlinecite{PhysRevB.84.245123}] and [\onlinecite{PhysRevB.88.214409}]). Ref. [\onlinecite{hurst}] studied the simplified case of a step-function out-of-plane skyrmion profile, and found that the charged skyrmion is only energetically favorable with an external magnetic field due to particle-hole symmetry at zero field. However, realistic skyrmion profiles are smooth and contain an in-plane region, which breaks particle-hole symmetry. Our results show that for realistic profiles with broken particle-hole symmetry, charged skyrmions may be favorable even at zero external field.
\par

The binding of charges to skyrmions is also seen in the quantum Hall ferromagnet, for instance at the filling factor $\nu=1$. In this state, the Coulomb repulsion favors the spontaneous polarization of electron spins and doping the $\nu=1$ ferromagnet leads to electrically-charged skyrmion textures \cite{sondhi,kane,girvin}. The quantum anomalous Hall state in intrinsic magnetic topological insulators differs from  the quantum Hall ferromagnet in several key aspects. First, its ferromagnetism comes from the ordering of local magnetic moments (with $\sim 5\mu_B$ Bohr magneton per Mn atom in the case of MnBi$_2$Te$_4$), rather than the spins of low-density charge carriers. Second, spin-momentum locking in the Dirac surface states of magnetic topological insulator leads to a strongly anisotropic magnetic response to the exchange field: an out-of-plane field opens a gap at charge neutrality, while an in-plane field does not. Therefore, the Chern-Simons effective theory for quantum Hall ferromagnets cannot be applied to describe the skyrmion physics in magnetic topological insulators. 

\textit{Very dilute limit}. Since the large-$S$ local magnetic moments are treated as classical, the skrymions and Dirac electrons  bound to them are immobile. When the density of added electrons is sufficiently low, the Coulomb repulsion is expected to drive a Wigner crystal phase of charged skyrmions. There are two distinct length scales in this phase: the lattice constants of the skrymion Wigner crystal (SWX), set by the density of doped electrons, and the skyrmion size, set by the exchange coupling and the spin susceptibility of Dirac electrons. The first length scale can far exceed the second. We emphasize that the SWX proposed here is driven by Wigner crystallization of charged skrymions. This mechanism is similar to the SWX phase  observed in the lightly doped $\nu=1$ quantum Hall ferromagnet \cite{brey,PhysRevLett.74.5112,skyrmiongraphene}, but different from the skrymion crystal phase  in helimagnets that is driven by DM interaction.

\section{Stripe phases}
\label{sec:stripe}

\begin{figure}
\centering
\includegraphics[width=\linewidth]{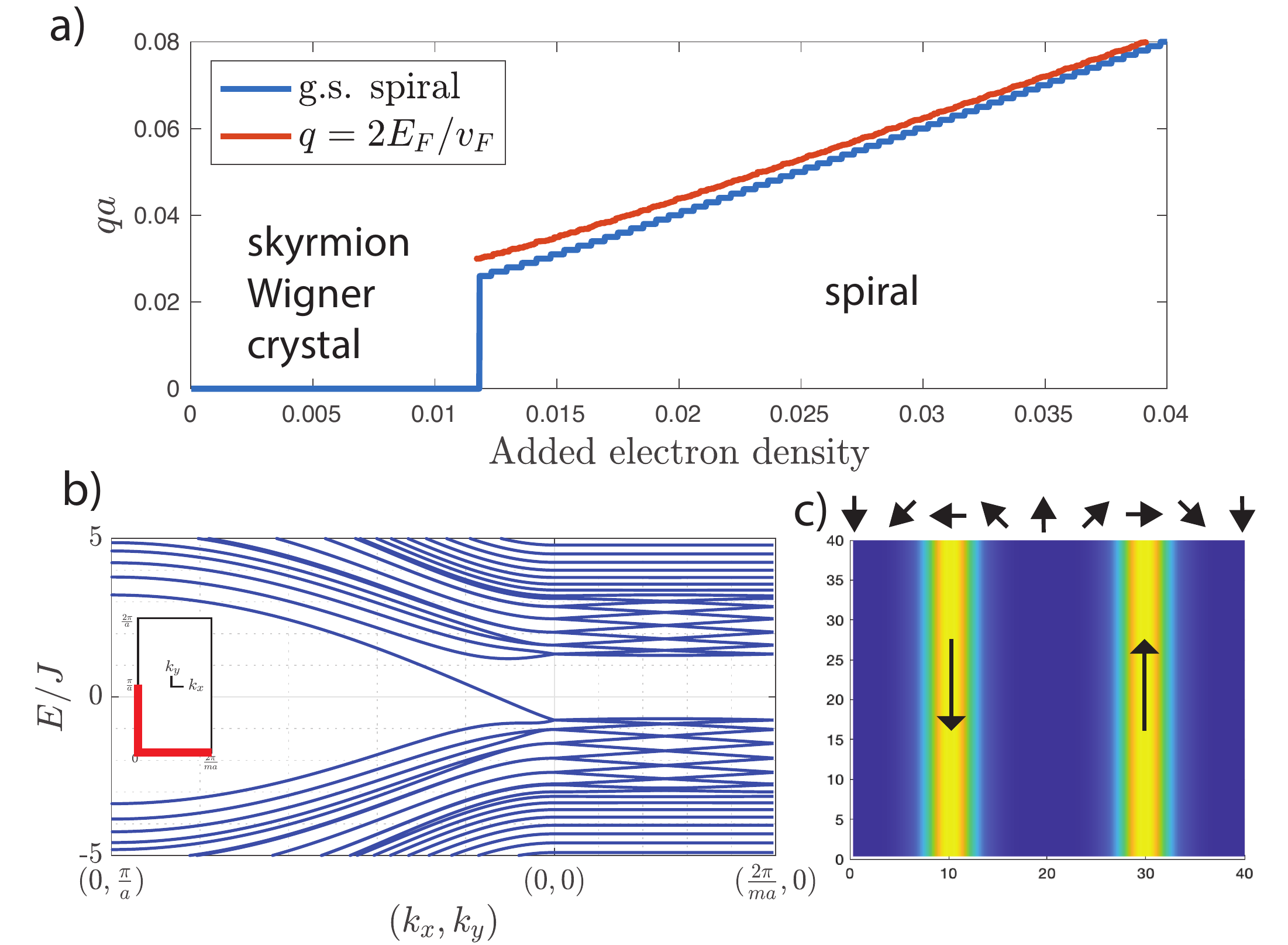}
\caption{(a) Ground state sinusoidal N\'eel spiral of $H_e+ H_{eS}$, characterized by wavevector $q$ in $x$ direction, as a function of doping. The slope $\sim 2k_F$ for the underlying dispersion. A $10^3\times500$ lattice with unit spacing and cutoff $k_{UV}= \pi/20a$ was used. (b) Spectrum along red line of inset mini BZ for Dirac electrons in a sinusoidal N\'eel spiral of wavelength $ma$ in the $x$ direction (here $m=20,J=0.5$). Conduction band states descend to form mid-gap bound states localized in the $x$ direction. Localization leads to nearly flat bands in the $k_x$ direction, with dispersion tuned by $J$. (c) Magnitude of spin up component of $E=0$ Dirac bound states in the presence of the spin spiral above. The velocities are indicated by long arrows.}
\label{fig:stripe}
\end{figure}

In this section we consider a finite electron density above charge neutrality. We work with the same tight-binding model (App. \ref{app:diag}) and the strong $J$ limit so that electrons actively dictate the ground state. We find the ferromagnetic state is unstable to a state with spatially modulated magnetization above a critical electron density, as shown in Fig. \ref{fig:stripe}a. Above the transition the preferred wavevector is $\sim 2k_F$.\par

Let us comment on the band structure of Dirac electrons in the presence of a spiral phase. We expect a network of 1d modes localized along the regions where spins are in-plane, as in Fig. \ref{fig:stripe}c. To see this, let us focus on a single such region, which we take along the $y$ axis, and approximate the spin texture in the vicinity as

\begin{equation}
\label{eq:qx}
\vec S(x,y) = (1,0,qx).
\end{equation}
The normalization $S^2=1$ is maintained to linear order in $x$. Solving the corresponding continuum equation
\begin{equation}
[v_F(-i \p_x \sig^y-k_y \sig^x)-J(\sig^x+qx \sig^z)]\psi = E\psi
\end{equation}
we obtain the bound state

\begin{equation}\label{eq:gaussian}
\psi_{k_y}(x,y) = C e^{ik_y y}e^{-\frac{J q x^2}{2v_F }} (1, -1)^T
\end{equation}
with linear dispersion $E= -\sgn(q)(v_F k_y+J)$. This state is Gaussian-localized in the $x$ direction over a length $\sim \sqrt{v_F/Jq}$. It follows that in the presence of a spiral texture $\vec S = (\cos qx, 0,\sin qx)$ well-localized, approximately degenerate chiral modes form when $J/v_F \gg q$, in which case Eq. \eqref{eq:qx} is a valid approximation. \par

In Fig. \ref{fig:stripe}b we plot the full spectrum of Dirac electrons in a N\'eel spiral texture to show that the linear branch of chiral modes descends into the gap from the conduction band, giving another picture for why a spiral state becomes favorable. Moreover, a spiral state in the appropriate limit would exhibit highly anisotropic conductance; the conductance would be much greater in the $y$ direction due to the 1d channels of bound states. Examples of such bound states are shown in Fig. \ref{fig:stripe}c.\par

Since the exchange coupling $J$ is small compared to the bandwidth of TI surface states (on the order of eV), the leading-order effect of $J$ on the spin degrees of freedom is to modify the spin-spin interaction through the RKKY mechanism. The effective Hamiltonian for the array of spins, after integrating out the electrons, is given by:
\begin{equation}\label{eq:chi}
H_{\text{eff}} = -\sum_{\bq ab} \chi_{\text{tot}}^{ab}(\bq)S^a_{-\bq}S^b_{\bq},
\end{equation}
where
\begin{equation}
\label{eq:chitot}
\chi_{\text{tot}}^{ab}(\bq) = J^2 \chi_{eS}^{ab}(\bq) + \chi_S^{ab}(\bq).
\end{equation}
The spin susceptibility $\chi_{\text{tot}}(\bq)$ is a $3\times3$ Hermitian matrix whose largest eigenvalue and corresponding eigenvector describe the wavevector and polarization of the ground-state spin texture. 
The first contribution is derived from second order perturbation theory in $Ja/v_F$ while the second term can be read off from $H_S$ when the Zeeman field is absent. Their explicit forms are given in Eq.s \eqref{eq:chiab} and \eqref{eq:chiS}. In the strong $J$ limit we drop $\chi_S$ and focus only on $\chi_{eS}$. The latter can can be written as $U^\dagger \tilde \chi U$ with $U$ a $3\times 3$ unitary effecting a $\pi/2$ rotation about $\hat z$ and

\begin{equation}
\label{eq:chiab}
\tilde \chi^{ab}(\bq) = \frac{-1}{2N} \sum_{\bk,s_1,s_2}\frac{f_{\bk,s_1}-f_{\bk+\bq,s_2}}{\xi_{\bk s_1}-\xi_{\bk+\bq,s_2}}F^a_{\bk s_1;\bk+\bq,s_2}F^b_{\bk+\bq,s_2;\bk s_1}
\end{equation}

with $\xi_{\bk s} = s|v_F|k-E_F$ and $f_{\bk s} = [1+e^{\beta \xi_{\bk s}}]^{-1}$ the Fermi distribution\cite{spinhelix}. The $F^a$ are given by

\begin{align}
F^x_{\bk_1s_1;\bk_2s_2} &= \frac{\sgn(v_F)}{2}(s_1 e^{i\theta_{\bk_1}}+s_2 e^{-i\theta_{\bk_2}})\\
F^y_{\bk_1s_1;\bk_2s_2}&= -i\frac{\sgn(v_F)}{2}(s_1e^{i\theta_{\bk_1}}-s_2e^{-i\theta_{\bk_2}})\\
F^z_{\bk_1s_1;\bk_2s_2}&=-\frac12(1-s_1s_2e^{i(\theta_{\bk_1}-\theta_{\bk_2})})
\end{align}
where $e^{i\theta_\bk} = (k_x+i k_y)/k$. \par

\begin{figure}
\centering
\includegraphics[width=.9\linewidth]{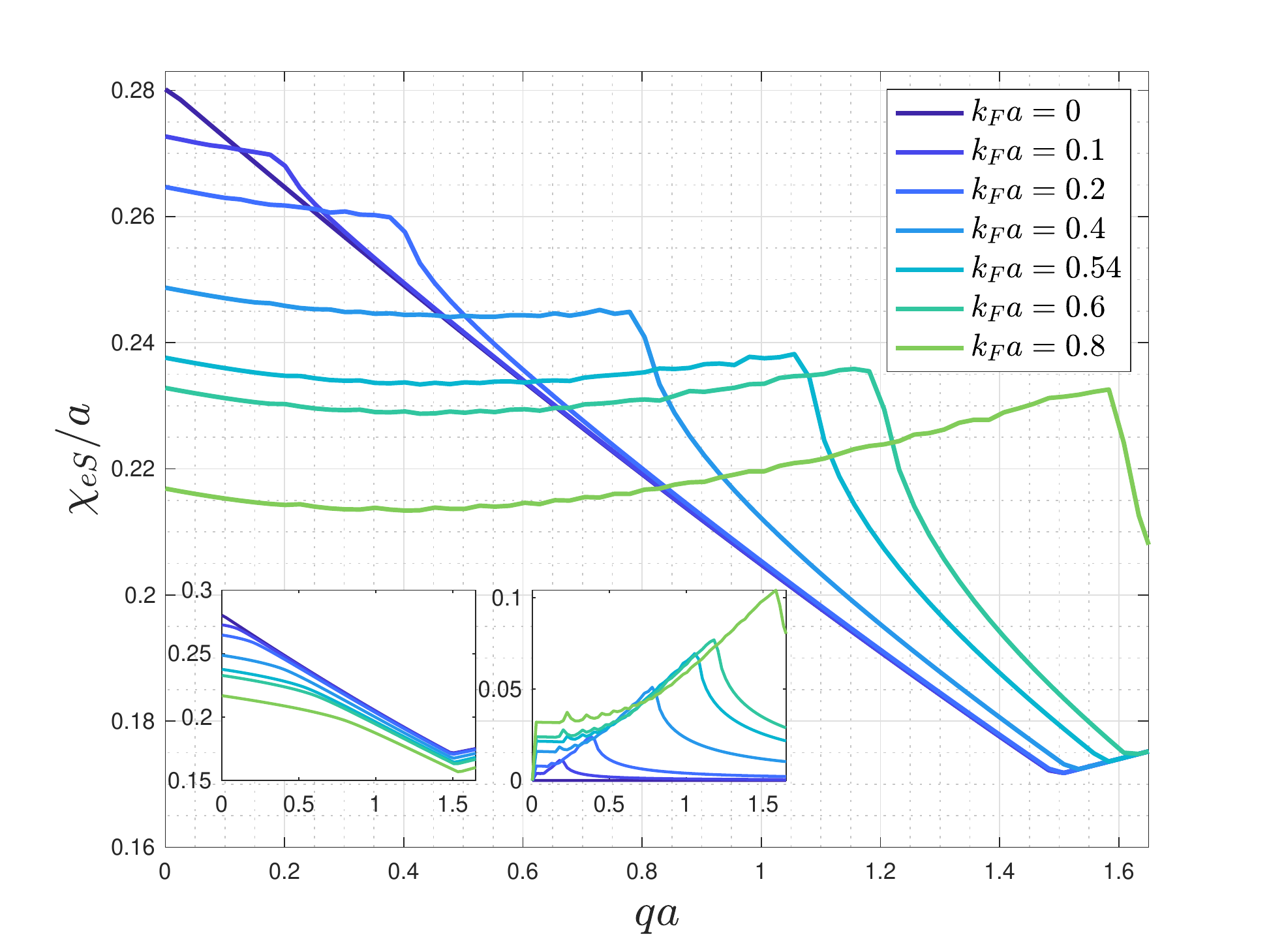}
\caption{The largest eigenvalue $\chi_{eS}(q)$ of the spin susceptibility defined in Eq. \eqref{eq:chi}. The dominant peak shifts discontinuously away from $q =0$ at a critical $k_Fa\approx 0.54$. The transition occurs at a point in the Brillouin zone where a Dirac cone is a good approximation to the dispersion of the $10^3\times500$ lattice model ($a=1$). Left and right insets display the inter- and intraband spin susceptibilities, respectively. The $2k_F$ peak is entirely due to the latter and the UV dependence is captured entirely in the former.}
\label{fig:rkky}
\end{figure}

In Fig. \ref{fig:rkky} we plot the dominant eigenvalue $\chi_{eS}(q)$ of the spin susceptibility varying the electron doping $k_F$ (taking $v_F=1$). A square lattice with lattice constant $a$ is used as a UV completion. The plot has a peak at $q=0$ for small doping which trades dominance with a peak at $q= 2k_F$ at some critical $k_F$. This corresponds to the transition from a uniform state to a $2k_F$ spiral, in accord with Fig. \ref{fig:stripe}a. The corresponding eigenvector describes a spiral texture with N\'eel wall-like rotation. For a non-Rashba spin-orbit coupling a Bloch wall-like rotation is favored. Fig. \ref{fig:rkky} is also consistent with a skyrmion state comprising a sum of multiple magnitude-$2k_F$ wavevectors, because $\chi_{eS}$ is radially symmetric up to isotropy-breaking terms originating from the underlying lattice.\par

We refer to Eq. \eqref{eq:chiab} with $s_1=s_2$ as the intraband susceptibility and with $s_1=-s_2$ as the interband susceptibility. The interband susceptibility is a UV divergent contribution (cut off by the lattice spacing) which occurs even when the chemical potential is zero and depends weakly on $q$. The intraband susceptibility is independent of UV cutoff and dominated by Fermi surface contributions. These are both plotted in Fig. \ref{fig:rkky} (inset). \par

To see how the skrymion Wigner crystal (SWX) phase fits into the picture presented in this section, we remark that the presence of two length scales in the SWX phase at low doping leads to multiple peaks in the Fourier transform of its magnetic structure, and the largest peak is at wavevector $\bq = 0$. Thus the closest single-wavevector approximation is the uniform state, consistent with Fig. \ref{fig:stripe}a and Fig. \ref{fig:rkky} in the low density regime. Higher order effects in the perturbative expansion in the exchange coupling $J$, higher harmonics in $\bq$ and the inclusion of Coulomb repulsion are needed to fully capture the SWX phase. \par

\textit{Intermediate regime.} We have previously worked at strong coupling, neglecting the effects of intrinsic magnetism. Now we allow the couplings of $H_{S}$ and $H_{eS}$ to be of the same order. The dominant eigenvalue of the total spin susceptibility (Eq. \eqref{eq:chitot}) determines the ground state properties of the system, where $\chi_{eS}$ was discussed in the previous section and $\chi_S$ can be read off from $H_S$ in the case of zero Zeeman field. Since a square lattice was used for $\chi_{eS}$ we use the same for $\chi_S$, although the only UV sensitivity will come from the interband susceptibility of $\chi_{eS}$, which affects the transition from $q=0$ to $q=2k_F$. Thus we have

\begin{equation}
H_S\left.\right|_{B=0}= -\sum_{ab} \chi_S^{ab}(\bq) S^a_{-\bq} S_\bq^b
\end{equation}
with $\chi_S(\bq)$ given by

\begin{subequations}
\label{eq:chiS}
\begin{align}
\chi^{xx}_S&=\chi^{yy}_S =A(\cos q_x a + \cos q_y a)\\
\chi^{zz}_S&=A(\cos q_x a + \cos q_y a)+K \\
\chi^{xz}_S&=(\chi^{zx}_S)^* = iD \sin q_x a \\
\chi^{yz}_S &= (\chi^{zy}_S)^* = iD \sin q_y a \\
\chi^{xy}_S&=\chi^{yx}_S=0.
\end{align}
\end{subequations}

\begin{figure}
\centering
\includegraphics[width=.9\linewidth]{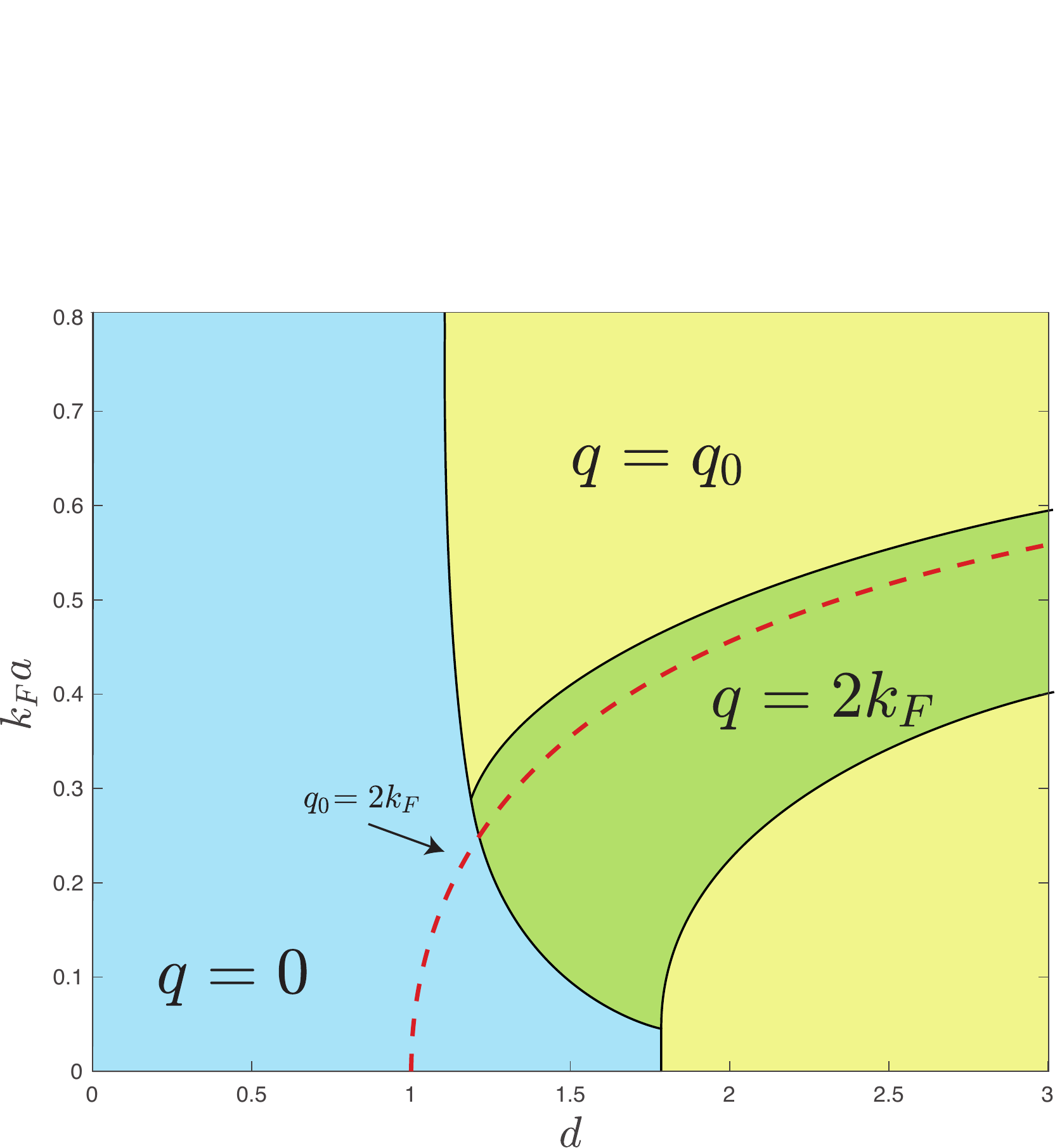}
\caption{Phase diagram describing Dirac electrons coupled to spins as the electron density and DM interaction are varied (within $(\Delta d, \Delta k_Fa) = (0.15,0.04)$) with $A=K=0.1, J=0.5$ and $a=1$ on a $100\times 100$ lattice. The spins exhibit stripe orders at wavevectors $q=0, 2k_F,$ and $q_0$ (defined in Eq. \eqref{eq:q0}), as determined by the spin susceptibility peak.}
\label{fig:pd}
\end{figure}

In terms of a rescaled coupling

\begin{equation}
d^2= 2D^2/AK
\end{equation}
the dominant eigenvalue of $\chi_S$ is peaked at $q=0$ for $d<1$ and at
\begin{equation}\label{eq:q0}
q_0 = \tan^{-1}(\sqrt{d^4-1}/\sqrt{2d^2A/K+1})/a
\end{equation}
for $d\geq 1$ if $\bq$ is taken along an axis of symmetry. The ground state of the system is a spin texture with wavevector $q=0,2k_F$, or $q_0$, determined by the peak of the dominant eigenvalue of $\chi_{\text{tot}}$. This competition yields the phase diagram shown in Fig. \ref{fig:pd}, plotted in the $(d, k_F)$ plane with $A=K=0.1, J=0.5$. \par

This phase diagram exhibits several interesting features. Increased electron doping expands the $q=q_0$ phase, thereby acting as an ``effective DM interaction," an effect noted in recent works \cite{skyrmionium1,dm1}. The line $q_0=2k_F$ broadens into the RKKY phase with increasing exchange coupling $J$. When $J$ vanishes, this phase vanishes and the $q=0$ and $q_0$ phases are separated by the transition at $d=1$. In the opposite limit $J\gg A,K$, the $q=q_0$ phase vanishes and we recover the transition in Fig. \ref{fig:rkky} as a function of $k_F$. We note that for weak exchange coupling $J$ one expects these phases to be stripe orders (single-$\bq$) rather than multiple-$\bq$ because for $J=0$ the stripe phase is known to be the ground state at zero external field.

\section{Discussion}
\label{sec:end}

We have investigated the phases of an array of magnetic moments coupled to Dirac electrons upon tuning the electron density, motivated by intrinsic magnetic TIs subject to electrostatic gating.

We found at very dilute densities electrons bind to magnetic skyrmions and we conjecture an SWX results; at higher densities the SWX gives way to spin spirals bound to chiral electron channels. The DM interaction resulting from broken inversion symmetry is essential for the formation of skrymions and stripes. Tuning the DM interaction and electron density reveals a phase diagram of stripe orders captured in Fig. \ref{fig:pd}. The interplay of real-space magnetic structures and topological Dirac electrons suggests the manipulation of magnetic domains and textures by electric currents and electrostatic gating is possible with low dissipation, an attractive prospect for skyrmion-based information devices. \par

An interesting future direction is the study of the effects of skyrmion and spiral phases on transport phenomena and quantum oscillations. In a skyrmion crystal phase the Berry flux attached to each skyrmion should lead to topological DOS oscillations which impact all physical observables \cite{PhysRevB.100.174411}. The emergent orbital magnetic field of the nontrivial spin textures leads to a topological contribution in the Hall resistivity $\rho_{xy}(B)$, observed in Mn$_2$CoAl thin films\cite{ludbrook}, Mn-doped Bi$_2$Te$_3$ quintuple layers\cite{PhysRevLett.119.176809}, and correlated oxide thin films\cite{Matsunoe1600304,vistoli} at low temperatures. However, anomalous features in Hall resistivity could originate from surface and bulk ferromagnetism instead of skyrmion physics \cite{molenkamp}. We leave the precise form of the contributions from topological magnetism to physical observables to future work. We hope that the magnetic phase transitions predicted in this paper may spark further interest in magnetic topological insulators.

\textit{Note added.} Recently two related works on spin textures in magnetic topological insulators have appeared \cite{paramekanti,chen}.\par

 \textit{Acknowledgments. } 
We thank Sungjoon Hong, Noah Yuan and Hiroki Isobe for participation and contribution in the early phase of this project. 
This work is supported by DOE Office of Basic Energy Sciences, Division of Materials Sciences and Engineering under Award DESC0018945. L.F. is partly supported by a Simons Investigator award from the Simons Foundation.

\appendix
\begin{widetext}

\section{Dirac electrons bound to a skyrmion}
\label{app:skyrm}
Here we find the bound state spectrum of Dirac electrons in the presence of a skyrmion spin texture. The spin texture is defined on a lattice, but it is convenient to take a continuum approximation

\begin{equation}
H_e = \int \psi^\dagger \mH \psi, \quad \mH = v_F(k_x\sig^y-k_y\sig^x)-J \vec S\cdot \vec \sigma.
\end{equation}
First we consider an idealized skyrmion of radius $b$, with

\begin{equation}\label{eq:A2}
S^z(r) = \begin{cases} -1& r< b\\
+1 & r>b
\end{cases}
\end{equation}
and $S^x = S^y = 0$. In radial coordinates and setting $v_F = J=1$ we may write

\begin{equation}
\mH = \sig^\theta (-i\p_r) +\frac1r\sig^r(-i\p_\theta)- S^z \sigma^z.
\end{equation}

It is convenient to define $\Lambda = \sqrt{r} e^{-\frac{i}{2}\sig^z(\theta+\frac{\pi}{2})}$ which satisfies

\begin{equation}\label{eq:L}
\Lambda^{-1}(\sig^r,\sig^\theta,\sig^z) \Lambda = (-\sig^y,\sig^x,\sig^z)
\end{equation}
and consider the isospectral Hamiltonian
\begin{equation}\label{eq:H'}
\mH' = \Lambda^{-1}\mH \Lambda = \sig^x(-i\p_r) -\frac1r\sig^y(-i\p_\theta)- S^z \sigma^z.
\end{equation}
Since $\mH'$ has no explicit $\theta$-dependence we replace $(-i\p_\theta)$ with a half-integer angular momentum $m$. Note that $m$ is a half-integer because $\Lambda$ changes the azimuthal boundary conditions from periodic to antiperiodic. We'd like to solve

\begin{equation}\label{eq:f}
[\sig^x (-i\p_r)- \frac{m}{r}\sig^y - S^z(r) \sig^z] \begin{pmatrix}
f_1 \\ f_2
\end{pmatrix} = E \begin{pmatrix}
f_1\\ f_2
\end{pmatrix}.
\end{equation}

Acting with $(-i\p_r \pm i \frac{m}{r})$ on the top and bottom row respectively we get
\begin{align}
- f_2'' + \frac{m(m-1)}{r^2} f_2 &= (E^2-1) f_2\\
- f_1'' + \frac{m(m+1)}{r^2} f_1 &= (E^2-1) f_1
\end{align}

away from $r=b$. These can be solved as
\begin{align}
f_2(x) &=\sqrt{x}(\alpha  J_{m-\frac12}(-i x) + \beta Y_{m-\frac12}(-i x))\\
f_1(x) &= -s\sqrt{x} \sqrt{\frac{1+sE}{1-sE}} (\alpha J_{m+\frac12}(-ix) + \beta Y_{m+\frac12}(-ix))
\end{align}
with $x= r\sqrt{1-E^2}$, $S^z(r) = s= \pm 1$, and $\alpha,\beta$ undetermined. There are several constraints on physical solutions, such as convergence of $\int d\theta dx\, x^2 |f|^2$ and conservation of probability current. Solutions in $r>b$ are restricted by convergence to have $\beta=-i\alpha$ and take the form

\begin{align}
f_2(x) &= \alpha' i^{m+\frac12} \sqrt{x} K_{m-\frac12}(x)\\
f_1(x) &=\alpha' i^{m-\frac12}\sqrt{\frac{1+E}{1-E}} \sqrt{x} K_{m+\frac12}(x).
\end{align}

Solutions in $r<b$ are restricted by conservation of probability current at the origin (all $m$) or convergence (for $|m|>1/2$) to have $\beta=0$ and take the form
\begin{align}
f_2(x) &= \alpha'' \sqrt{x} (-i)^{m-\frac12}I_{m-\frac12}(x) \\
f_1(x) &= \alpha'' \sqrt{\frac{1-E}{1+E}} \sqrt{x} (-i)^{m+\frac12} I_{m+\frac12}(x)
\end{align}
using $J_\nu(-ix)=(-i)^\nu I_\nu(x)$. Continuity at $x_0 = b\sqrt{1-E^2}$ fixes $\alpha'$ and the energy:

\begin{equation}
1 = \frac{1+E}{1-E} \frac{I_{m+\frac12}(x_0) K_{m-\frac12}(x_0)}{I_{m-\frac12}(x_0) K_{m+\frac12}(x_0)}.
\end{equation}

This is a transcendental equation we can solve to get the bound state spectrum for any angular momentum $m$ and radius $b$. The bound states have $E \approx m/b$ with great accuracy for large $b$ and some deviation at small $b$. They exist only for $|m|\lesssim b$. Importantly, they are localized near the radial domain wall $r=b$. \par

Indeed, one could have approximated this spectrum to great accuracy with the ansatz of a $-1$ eigenspinor of $\sig^y$. Then one finds $f_1,f_2 \sim \exp(-S^z(r) r)$ from Eq. \eqref{eq:f}, implying $f_1,f_2$ are exponentially localized near $b$. This allows the approximation $m/r\approx m/b$ to reproduce the bound state spectrum. \par
This approximation is useful when we consider what happens upon introducing a small radial in-plane region to the spin texture, i.e. a region $(b,b+\delta)$ in which $S^z=0$ and $S^r=\pm 1$. Keeping in mind our choice of $\Lambda$ in Eq. \eqref{eq:L}, we observe that this is captured by the perturbation $\theta(r-b)\theta(b+\delta-r) (\mp \sig^y+\sig^z)$. Since $(f_1,f_2)^T$ is an approximate $-1$ eigenspinor of $\sig^y$ we find a uniform $m$-independent spectral shift in first order perturbation theory, breaking particle-hole symmetry. To leading order in $\delta$,
\begin{equation}\label{eq:fr0}
\Delta E \approx \mp\delta \frac{4b^2 e^{-2b}}{(4b^2+2)\cosh2b-4b\sinh2b-1}.
\end{equation}

We confirmed the shift by full analytical solution. \par
For comparison, we investigate the bound states of Schr\"odinger electrons in the same idealized skyrmion texture Eq. \eqref{eq:A2} with Hamiltonian
\begin{equation}
\mH = \frac{1}{2\mu}(k_x^2+k_y^2)\sig^z - J \vec S\cdot \vec \sig.
\end{equation}
In radial coordinates, replacing $-i\p_\theta$ with angular momentum $m$ and taking an $s$-eigenstate of $\sig^3$, the Schr\"odinger equation becomes
\begin{equation}
\frac{1}{2\mu}[-\p_r^2-\frac1r \p_r + \frac{m^2}{r^2} - J] f= sE f,\qquad (r<b)
\end{equation}
where $f$ is the nonzero component of the spinor. The solution is oscillatory and peaked near the origin rather than at $b$. For $r>b$ the solution decays. This sharply contrasts with Dirac bound states, which are exponentially peaked at $b$. An energetics analysis for a Schr\"odinger electron with spin-orbit coupling in the presence of a skyrmion texture was carried out in Ref. [\onlinecite{brey}].

\section{Tight-binding model for numerics}
\label{app:diag}

Numerical calculations were performed using a tight-binding approximation to Eq. \eqref{eq:dirac}, with an additional term to avoid fermion doubling. On a square lattice of lattice constant $a$, we took

\begin{equation}
H_{e}+H_{eS} = v_F\sum_{\br} \frac{i}{2a}\hat c_\br^\dagger (\sig^y(\hat c_{\br+\hat \bx}-\hat c_{\br-\hat \bx})-\sig^x(\hat c_{\br+\hat{ \mathbf{y}}}-\hat c_{\br-\hat{ \mathbf{y}}}))-v_F\sum_{\br,\mathbf{e}=\hat\bx,\hat{\mathbf{y}}} \frac{1}{2a}\hat c_{\br}^\dagger \sig^z(2\hat c_{\br}-\hat c_{\br+\mathbf{e}}-\hat c_{\br-\mathbf{e}})-J\sum_{\br} \hat c_{\br}^\dagger \vec S_\br \cdot \vec \sigma \hat c_\br.
\end{equation}

We set $a=v_F=1$ throughout. In Figs. \ref{fig:sk} and \ref{fig:sk2} we solved the spectrum of $H_e+H_{eS}$ above on a 50$\times$50 lattice in the presence of the skyrmion textures shown. We plotted the bound states and the momentum cutoff $k_{UV}$ below which the skyrmions are favorable to the uniform state. In the interacting case the momentum cutoff is implemented by keeping states within a certain range determined by $k_{UV}$ around charge neutrality. A similar procedure was used in Fig.s \ref{fig:rkky} and \ref{fig:stripe} although $y$-translation invariance was leveraged to use a 1000$\times$500 lattice and no $k_{UV}$ was used in Fig. \ref{fig:rkky}.

\end{widetext}

\bibliography{magtop}
\bibliographystyle{prsty}

\end{document}